\documentclass[journal]{IEEEtran}

\usepackage{booktabs}
\usepackage{tabularx}

\usepackage{amssymb}
\usepackage{amsbsy}
\usepackage{amsmath}
\usepackage{bm}
\usepackage{verbatim}
\usepackage{cite}
\usepackage{mathrsfs}
\usepackage{amsfonts}
\usepackage{graphicx}
\usepackage[tight,footnotesize]{subfigure}
\usepackage[10pt]{moresize}
\usepackage{array}
\usepackage{color}
\usepackage{epsfig}
\usepackage{stfloats}
\usepackage{balance}

\usepackage{subfigure}

\usepackage[noend]{algpseudocode}

\usepackage{algorithmicx,algorithm}

\usepackage{cite}

\usepackage{setspace}
\usepackage{cases}

\usepackage{graphicx}
\usepackage{epstopdf}
\usepackage{multirow}
\usepackage{extarrows}
\newcommand{\subparagraph}{}
\usepackage{titlesec}

\titlespacing{\section}{0pt}{2ex plus .0ex minus .0ex}{1ex plus .0ex}
\titlespacing{\subsection}{0pt}{1ex plus .0ex minus .0ex}{0.5ex plus 0.0ex}
\titlespacing{\subsubsection}{0pt}{0.0ex plus .0ex minus .0ex}{0.0ex plus .0ex}
\usepackage{amsmath} 
\allowdisplaybreaks[4]

\ifCLASSINFOpdf

\else

\fi

\begin{document}

\title{Microwave QR Code: An IRS-Based Solution}
%
\author{{Sai~Xu,~\IEEEmembership{Member,~IEEE,}
~Yanan Du, ~\IEEEmembership{Graduate Student Member,~IEEE,}\\
~Jiliang Zhang, ~\IEEEmembership{Senior~Member,~IEEE,}
and~Jie~Zhang,~\IEEEmembership{Senior~Member,~IEEE} 

\thanks{Sai~Xu (e-mail: \texttt{xusai@nwpu.edu.cn}) is with the School of Cybersecurity, Northwestern Polytechnical University, Xi'an, Shaanxi, 710072, China, and also the Department of Electronic and Electrical Engineering, University of Sheffield, Sheffield, S1 4ET, UK. Yanan~Du (e-mail: \texttt{ynduyndu@163.com}) is with the School of Cybersecurity, Northwestern Polytechnical University, Xi'an, Shaanxi, 710072, China. Jiliang~Zhang (e-mail: \texttt{jiliang.zhang@sheffield.ac.uk}) and Jie~Zhang (e-mail: \texttt{jie.zhang@sheffield.ac.uk}) are with the Department of Electronic and Electrical Engineering, University of Sheffield, Sheffield, S1 4ET, UK. (\emph{corresponding author: Jiliang~Zhang})}}}

%
\maketitle
\begin{abstract}
This letter proposes to employ intelligent reflecting surface (IRS) as an information media to display a microwave quick response (QR) code for Internet-of-Things applications. To be specific, an IRS is used to form a dynamic bitmap image thanks to its tunable elements. With a QR code shown on the IRS, the transmitting and receiving antenna arrays are jointly designed to scan it by radiating electromagnetic wave as well as receiving and detecting the reflected signal. Based on such an idea, an IRS enabled information and communication system is modelled. Accordingly, some fundamental systematic operating mechanisms are investigated, involving derivation of average bit error probability for signal modulation, QR code implementation on an IRS, transmission design, detection, etc. The simulations are performed to show the achievable communication performance of system and confirm the feasibility of IRS-based microwave QR code.
\end{abstract}
%
\begin{IEEEkeywords}
Quick response (QR) code, intelligent reflecting surface, reconfigurable intelligent surface, modulation, detection.
\end{IEEEkeywords}
%
%
\IEEEpeerreviewmaketitle
%
\section{Introduction}
\IEEEPARstart{A}LTHOUGH intelligent reflecting surface (IRS) has been widely investigated~\cite{GongToward}, almost all of the studies treat it as a pure reflection device. To be specific, IRS is a two-dimensional metasurface, on which large numbers of electromagnetic (EM) sensitive elements are printed~\cite{Wu2020Towards}. Compared to active antennas, the major distinction of IRS lies in that none of transmitting radio frequency (RF) chains is equipped~\cite{Xu2021Envisioning}. When illuminating an IRS, the amplitude and phase of incident signal can be changed by adjusting the reflection coefficient of elements. Owing to such an appealing ability, IRS can be used to construct a smart active environment to enhance or degrade signal reception~\cite{Renzo2020Smart}. Typically, IRS has been considered in multi-user communication~\cite{Zhang2021MultiUser}, cognitive radio~\cite{Xu2021Intelligent}, physical layer security~\cite{Xu2021Intelligent}, device-to-device communication~\cite{Mao2021Intelligent}, multi-cell communication~\cite{Khan2020Centralized}, non-orthogonal multiple access~\cite{Ding2020Simple}, ect. Up to now, however, IRS has not yet been considered as an information carrier, or more accurately an EM image displayer, to map the information to the complex reflection coefficient of its elements except for the similar concept proposed in our recent work~\cite{XuarXivIRSbackscatter}.\par
A supremely excellent case for the IRS-based EM image displayer is to show a microwave quick response (QR) code, which has many advantages, such as low power consumption, easy execution, simple collaboration mode with transceivers, sound and systematic QR technology base, modulation and coding easy to implement, insensitivity to weather and obstruction, long-distance and reliable communication, etc. Compared to conventional antenna-based communication, microwave QR code can lighten terminal devices, as well-known edge computing~\cite{Mach2017Mobile} and precoding~\cite{Agiwal2016Next} have done. Moreover, passive beamforming is not required in comparison with existing IRS backscatter~\cite{Xu2022Intelligent,Tang2020MIMO}.
Therefore, microwave QR code is an ideal technology for Internet-of-Things and low-power communication scenarios, such as wireless sensor networks, smart grid, logistics management, maritime affairs, etc.\par
Motivated by these distinctive characteristics, this letter will focus on IRS-based microwave QR code, which provides a transformative means to implement wireless communication. The contributions of this letter are summarized as follows. 1) An concept of microwave QR code is proposed based on our recent work~\cite{XuarXivIRSbackscatter}, which can be executed on an IRS, but not limited to it. The proposed system framework also applies to other cases of coded image, in addition to QR code. 2) Fundamental systematic operating mechanisms are preliminarily presented, including signal modulation, coding, transmission design, detection, etc. 3) Simulations are performed to show the feasibility of the proposed system and schemes, which are able to provide useful guidelines to inspire future research.
\section{System Model}
\begin{figure}
\centering
\includegraphics[width= 3.5 in]{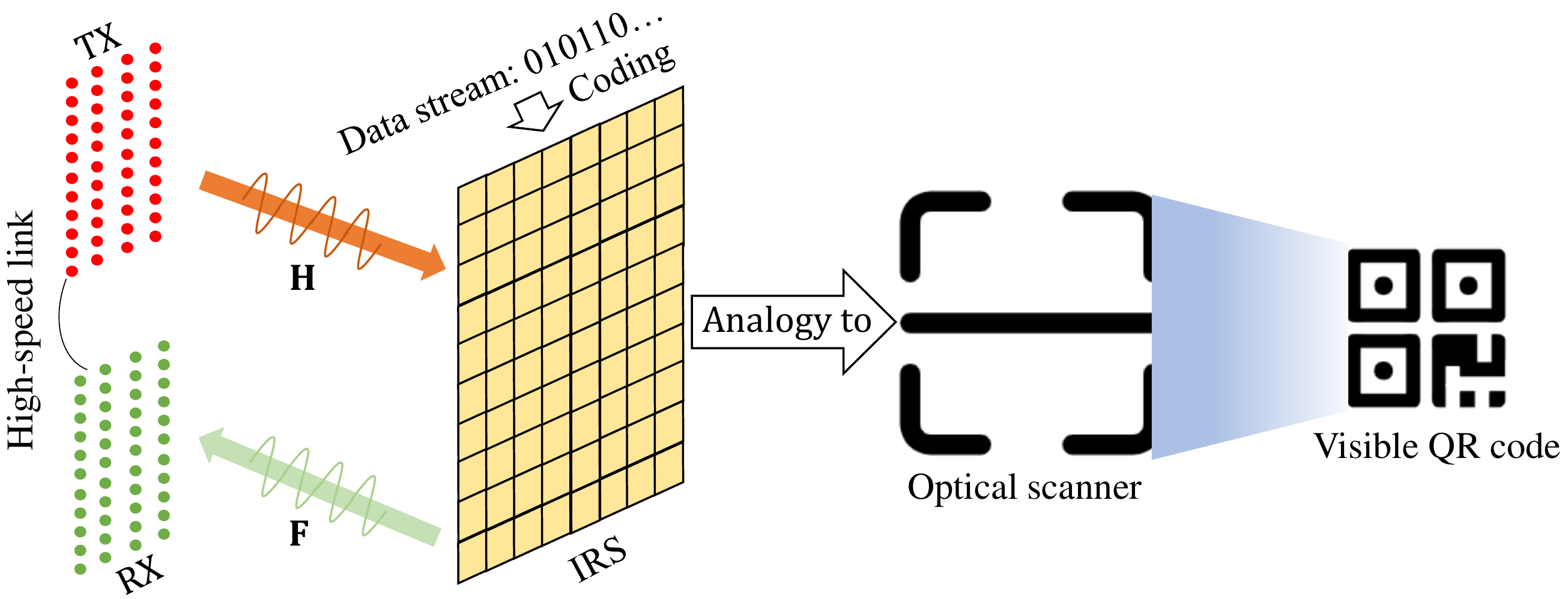}
\caption{An illustration of the proposed IRS enabled information and communication system, where an IRS acts as an element image displayer to map the information to the complex reflection coefficient of its elements. }
\label{Fig1}
\end{figure}

Consider an IRS enabled information and communication system as shown in Fig. \ref{Fig1}, where an IRS builds up an element image carrying data information instead of a pure reflector for radio signal, while the transmitting array antennas (TXs) radiate EM wave towards the IRS and the receiving array antennas (RXs) detect the echo signal for data extraction. Between the TXs and the RXs, no self-interference is assumed. Moreover, the TXs and the RXs may be deployed centrally or separately and can collaborate to read the information on the IRS. The relationship between the TXs/RXs and the IRS is in analogy with that between an optical scanner and a visible QR code.\par
Let $N_\text{t}$, $N_\text{r}$ and $L$ denote the numbers of the TXs, the RXs and the elements of IRS, respectively. It is assumed that all channel states are perfectly known with quasi-static frequency selective fading. $\textbf{H} \in {\mathbb{C}^{L \times N_\text{t}}}$ and $\textbf{F} \in {\mathbb{C}^{N_\text{r} \times L}}$ denote the channel gain matrices from the TXs to the IRS and from the IRS to the RXs, respectively. When the TXs radiate EM wave bearing no information, each IRS can passively modulate its information to the reflected signal with time-varying reflection coefficient of each element. Mathematically, the modulation process at the IRS is modelled as
\begin{align}
\textbf{F} \mathbf{\Theta}  \textbf{H} \textbf{w}
= \textbf{F} \text{diag} \{ \textbf{H} \textbf{w} \} \boldsymbol{\theta}, \label{eq:1}
\end{align}
where $\textbf{w}$ is the beamforming vector at the TXs.  $\boldsymbol{\Theta}$ and $\boldsymbol{\theta}$ are respectively the diagonal matrix of reflection coefficient  and the corresponding vector, with $\boldsymbol{\Theta} = \text{diag} \{ \boldsymbol{\theta} \}$. $\boldsymbol{\theta}$ is an information vector rather than a passive beamforming vector, where each element is not more than one. Therefore, $\left[ \boldsymbol{\theta}  \boldsymbol{\theta}^H \right]_{l,l}\leq 1 $ holds, where $\left[\cdot \right]_{l,l}$ denotes the $l$-th diagonal element of matrix. Based on this model, the received signal at the RXs is given by
\begin{align}
\textbf{y}  =  \textbf{F} \text{diag} \{ \textbf{H} \textbf{w} \} \boldsymbol{\theta} +  \textbf{z}, \label{eq:2}
\end{align}
where $ \textbf{z}$ is white Gaussian noise vector with $ \textbf{z} \sim (\textbf{0}, \sigma^2 \textbf{I}) $.\par
The performance of the considered IRS enabled information and communication system is highly dependent on the amplitude and phase of each element. Moreover, the number of RXs has a huge impact on the detection of the echo signal and information acquisition. In the following sections, we will investigate two related agendas for the considered system, including average bit error probability (ABEP) of signal modulation and IRS enabled QR code communication.
\section{Derivation of ABEP}
This section will compute the average symbol error probability (ASEP) and the ABEP for the considered IRS enabled information and communication system according to the information theory. Based on this, the ABEP of phase shift keying (PSK) as a typical modulation example is given, considering that the amplitude adjustment of elements at the IRS has a high control complexity. \par
Specifically, let $\textbf{V} = \textbf{F} \text{diag} \{ \textbf{H} \textbf{w} \}$ and $\textbf{U} = [\textbf{V}^H \textbf{V}]^{-1} \textbf{V}^H $. Based on \eqref{eq:2}, the output of the received signal at the RXs is given by
\begin{align}
\textbf{y}'  =   \boldsymbol{\theta} + \textbf{z}', \label{eq:3}
\end{align}
where $\textbf{y}' = \textbf{U} \textbf{y}$ and $\textbf{z}' = \textbf{U} \textbf{z}$. For the $l$-th element of IRS, the corresponding output of the received signal is given by
\begin{align}
\textbf{y}'(l)  =   \boldsymbol{\theta}(l) + \textbf{z}'(l). \label{eq:4}
\end{align}
It is assumed that the amplitude and phase of each element at the IRS construct an combined signal constellation diagram of size $M$, denoted by $\mathcal{X} = \{x_1, x_2, \cdots, x_M \}$, with all constellation points having the same occurrence probability. Therefore, the maximum likelihood (ML) estimate of $\boldsymbol{\theta}(l) $ is given by
\begin{align}
\hat{\boldsymbol{\theta}} (l) = \arg\min_{x_m \in \mathcal{X}} |\textbf{y}'(l) - x_m|. \label{eq:5}
\end{align}
Then, the closed-form expression of ASEP for the $l$-th element of IRS is given by
\begin{align}
\text{P}_\text{ASER} (l)
 =&  \sum_{m=1}^M \sum_{\hat{m}=1, \hat{m}\neq m}^M \frac{\text{Pr} \left\{ \hat{\boldsymbol{\theta}}(l) = x_{\hat{m}} | \boldsymbol{\theta}(l) = x_m \right\}}{M}. \label{eq:6}
\end{align}
Based on \eqref{eq:6}, the closed-form expression of ABEP for the $l$-th element of IRS can be obtained. When each element of IRS switches only between two states, the low-complexity modulation of BPSK is an excellent choice. For BPSK, the ABEP for the $l$-th element of IRS is given by
\begin{align}
\text{P}_\text{ABEP, BPSK}(l)
 = Q \left( \sqrt{ \frac{2}{ \textbf{C}_{ll}} } \right).
\end{align}
When more phase states can be adjusted for each element, more information is contained at a fixed-size IRS. For QPSK, the ABEP for the $l$-th element of IRS is approximately given by
\begin{align}
\text{P}_\text{ABEP, QPSK} (l)
\approx
Q \left( \sqrt{ \frac{1-\cos(\pi/2)}{\textbf{C}_{ll}}}\right).
\end{align}
When the number of phase states is greater than four, the ABEP for the $l$-the element of IRS is approximately given by
\begin{align}
\text{P}_\text{ABEP, \emph{M}-PSK} (l)
&\approx
\frac{
2}{\log_2 M} \Bigg[ Q \left( \sqrt{ \frac{1-\cos(2\pi/M)}{\textbf{C}_{ll}}}\right)   \nonumber\\
&  + Q \left(\sqrt{ \frac{1-\cos(4\pi/M)}{\textbf{C}_{ll}}}\right) \Bigg]
, ~ M>4,
\end{align}
where $\textbf{C}_{ll}$ is the $(l,l)$-th element of $\textbf{C} = \sigma^2 \textbf{U} \textbf{U}^H$.
\section{IRS enabled QR Code Communication}
%
%
%
\begin{figure}
\centering
\includegraphics[width= 3.2 in]{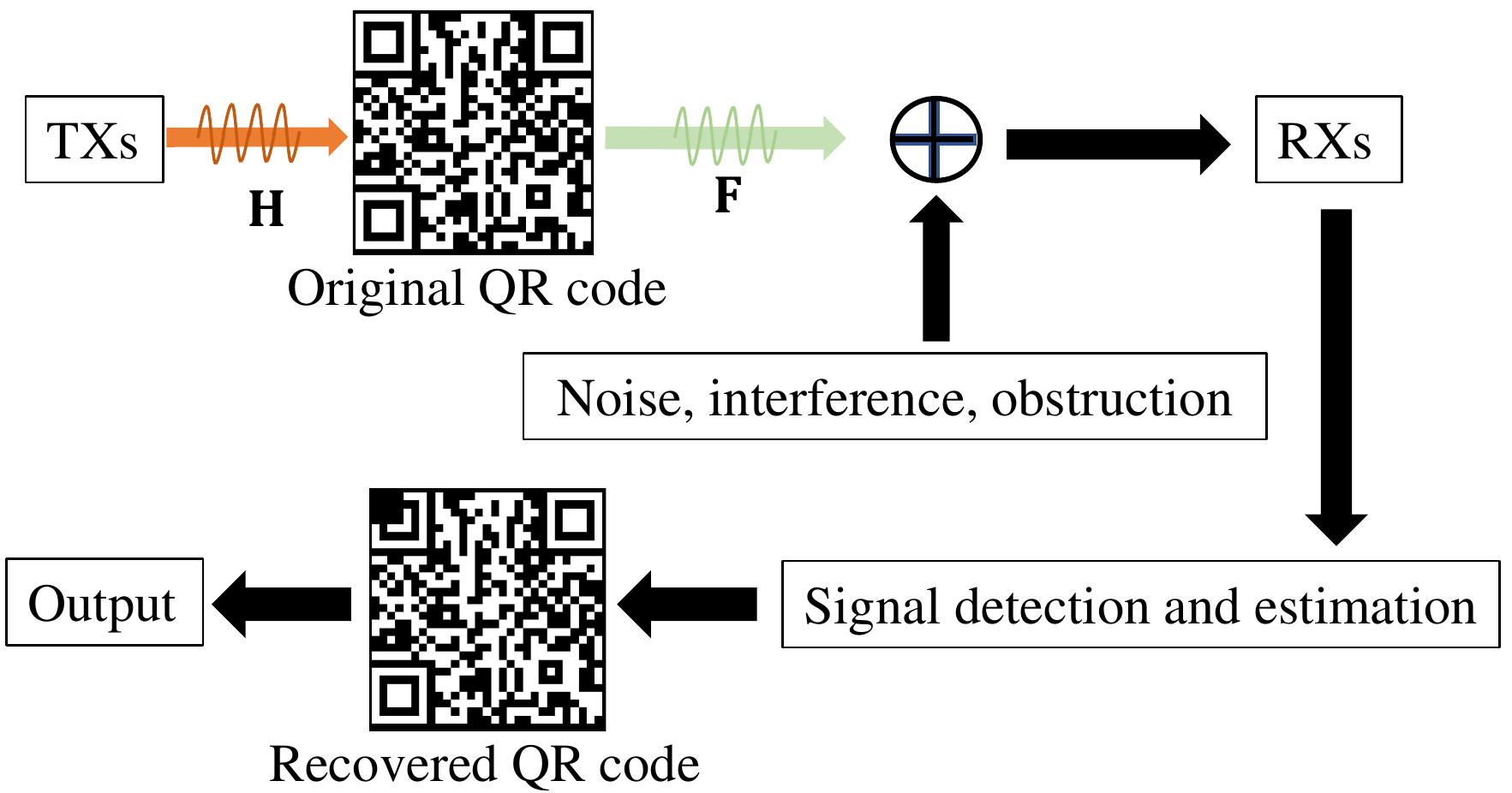}
\caption{An illustration of the principle of IRS enabled QR code communication. }
\label{Fig2}
\end{figure}
\begin{figure}
\centering
\includegraphics[width= 3.2 in]{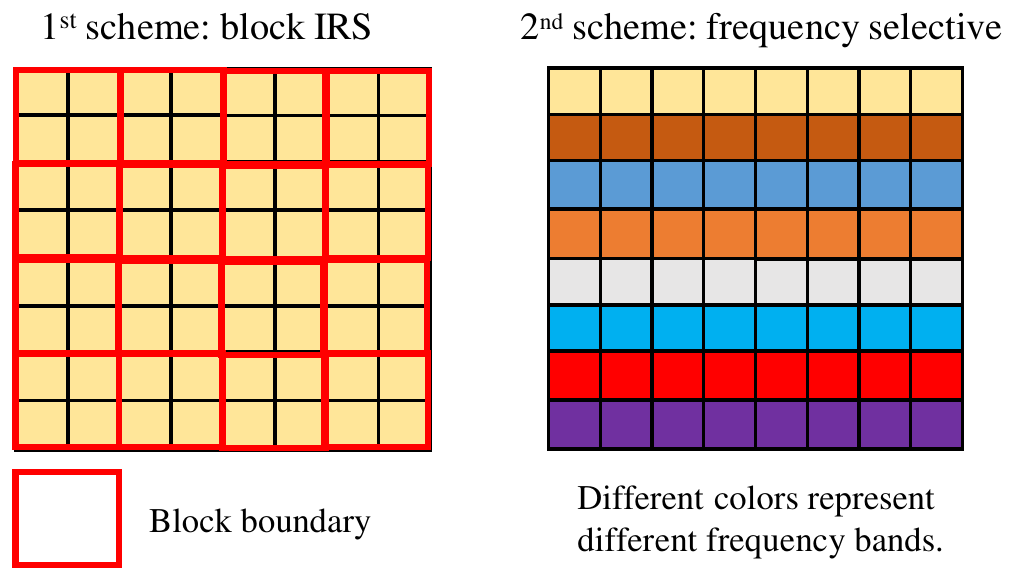}
\caption{Two schemes to cope with the insufficient RXs. }
\label{Fig3}
\end{figure}
This section will discuss how an IRS as a QR code displayer interacts with the TXs/RXs, involving QR code implementation, transmission design, detection, etc. The operating principle is depicted in Fig. \ref{Fig2}. Considering that there are many versions for QR code to support different languages, numbers, letters or their mixture and that QR coding is an independent field of study, the QR coding design is beyond the scope of this letter.\par
At the TXs, channel estimation and beamforming design have a huge effect on the communication performance of system. Since IRS acts as an information carrier with each element having the identical constellation diagram, $\mathbf{\Theta} = \textbf{I}$ is set as the precondition for channel estimation and beamforming. Under such a precondition, it is easy to estimate the cascaded channel $\textbf{F} \mathbf{\Theta} \textbf{H}$, that is $\textbf{F} \textbf{H}$. Based on the channel state information of $\textbf{F} \textbf{H}$, a straightforward beamforming scheme is adopted, where $\textbf{w}$ is designed as the eigenvalue corresponding to the largest eigenvalue of the matrix $\textbf{H}^H \textbf{F}^H \textbf{F} \textbf{H}$.\par
At the RXs, the number of antennas may be far smaller than that of IRS elements in practice, which results in the difficulty in the signal detection. To address this issue, two schemes may be considered, as illustrated in Fig. \ref{Fig3}. In the first scheme, an IRS is equally divided into $N_\text{r}$ blocks while all elements in a block have the same reflection coefficient~\cite{Xu2022Intelligent}. As a result, the QR modules are reduced into $N_\text{r}$, each of which can reflect more power towards the RXs. If the number of QR modules is insufficient to bear a complete message, QR codes carried by multiple frames on the IRS are merged into one for information transfer. In the second scheme, frequency division multiplexing is employed. To be specific, an IRS is made up of multiple groups of frequency selective elements for interference elimination and each group is able to reflect EM wave in a specific frequency band. Based on this scheme, the signal reflected by different groups of elements can be detected by the RXs at the same time.\par
Relying on different modulation schemes, each element or IRS block is used as one or multiple QR modules. For example, when BPSK is adopted, one element or IRS block represents one QR module; when 16-PSK is adopted, one element or IRS block represents four QR modules. When the EM wave from the TXs impinges the IRS, the information carried by each element is modulated into the reflected signal. Then, the signal is received, detected and estimated at the RXs. Due to noise, interference and obstruction, the recovered QR code may be impaired. Even so, the information can still be correctly decoded under a certain impaired level.
%
%
\section{Numerical Results}
\begin{figure*}[htb]
\begin{minipage}[t]{0.33\linewidth}
\centering
\includegraphics[width=1\textwidth]{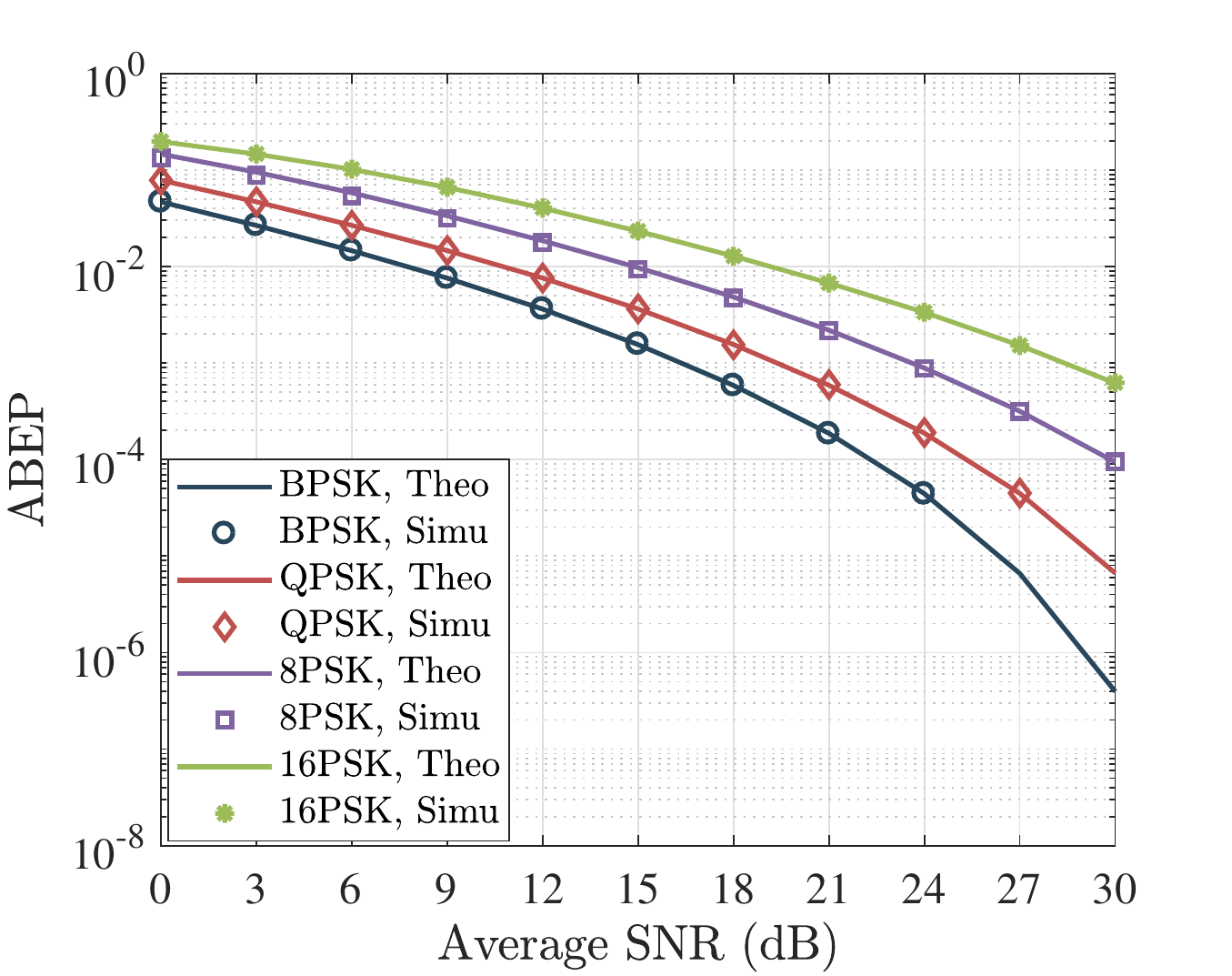}
\caption{ABEP \emph{vs.} average SINR.}
\label{Fig4}
\end{minipage}
\begin{minipage}[t]{0.33\linewidth}
\centering
\includegraphics[width=1\textwidth]{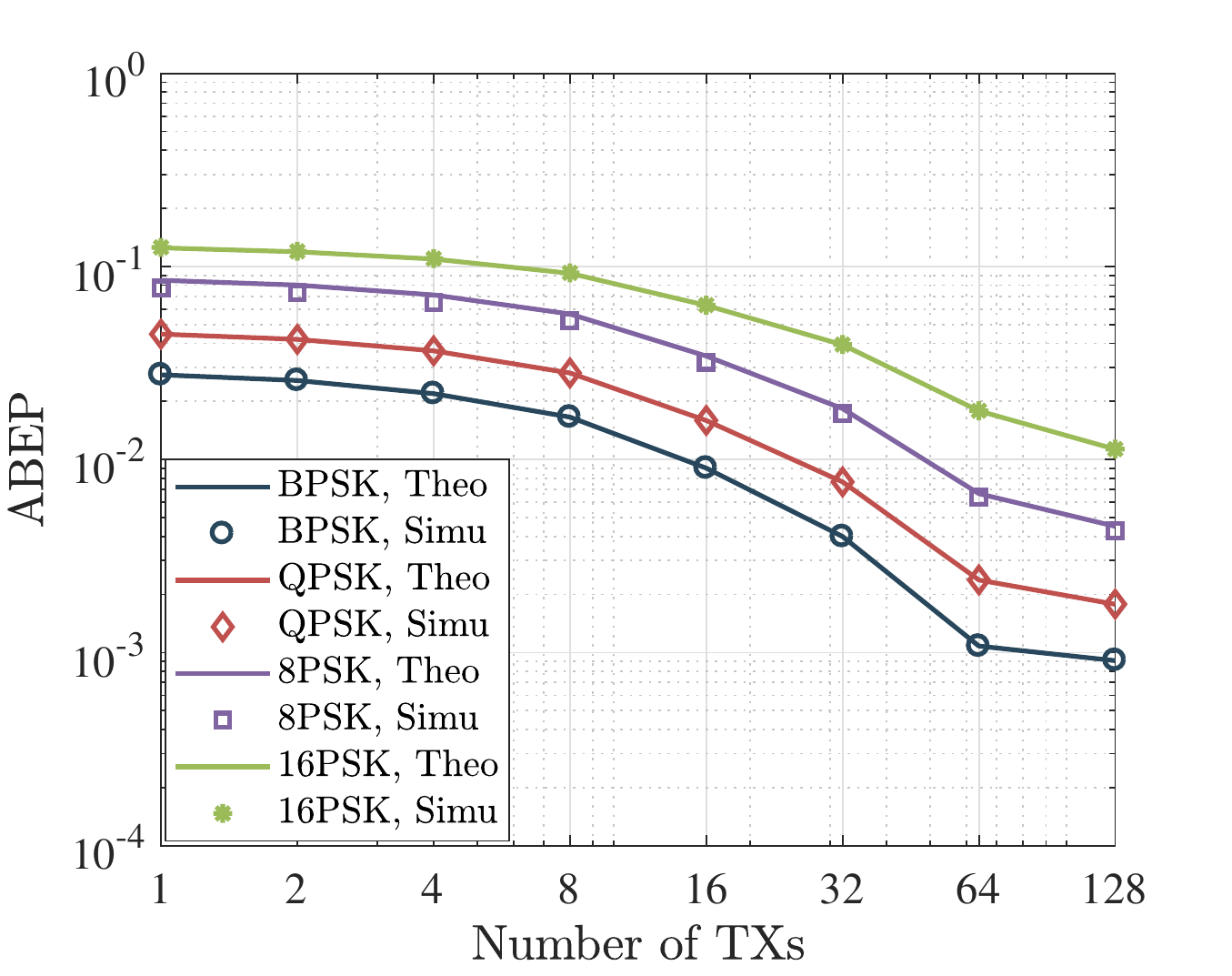}
\caption{ABEP \emph{vs.} number of TXs.}
\label{Fig5}
\end{minipage}
\begin{minipage}[t]{0.33\linewidth}
\centering
\includegraphics[width=1\textwidth]{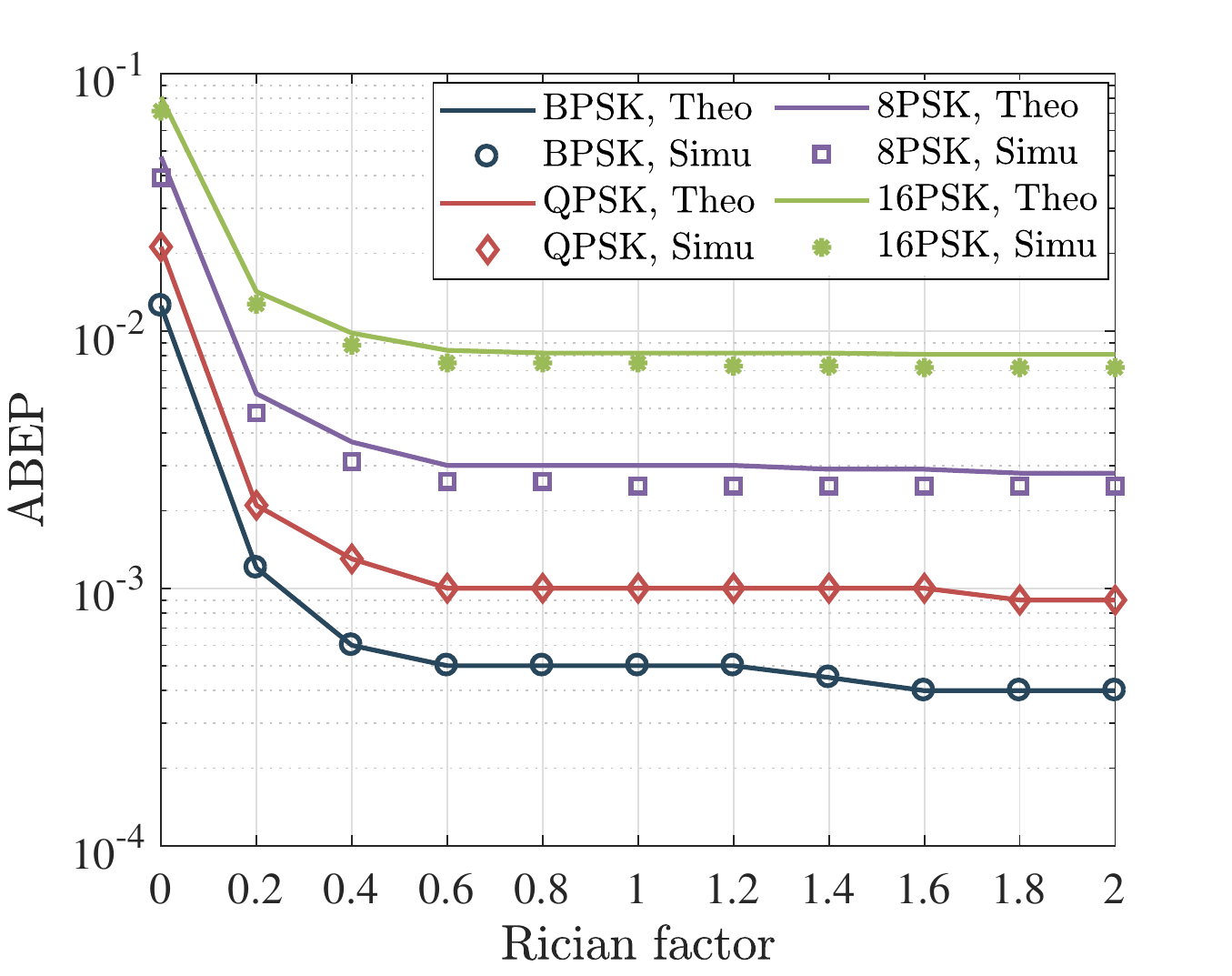}
\caption{ABEP \emph{vs.} Rician factor.}
\label{Fig6}
\end{minipage}
\end{figure*}
This section will evaluate the feasibility and the achievable performance of the proposed IRS enabled information and communication system by numerical simulations. In simulations, the channels $\textbf{H}$ and $\textbf{F}$ are randomly generated from Rician channel distribution with the Rician factor $\kappa$, where the path loss at the transmission distance $d$ is given by $\text{PL} = \text{PL}_0 - 25 \lg \left( {d}/{d_0} \right) $ dB with $\text{PL}_0$ = -30 dB at the reference distance $d_0$ = 1m. The legend \texttt{PSK} represents the schemes of PSK with multiple modulation orders considered, while \texttt{Theo} and \texttt{Simu} denote the theoretical and simulation results, respectively. Some constant parameters are set as: the distance from the TXs/RXs to the IRS $d_\text{I}$ = 50 m, the noise temperature $T$ = 300 K, and the noise bandwidth $B$ = 1 MHz.\par
Figs. \ref{Fig4}, \ref{Fig5} and \ref{Fig6} depict how the ABEPs of multiple modulation schemes of PSK depend on the average signal-to-noise-ratio (SNR), the number of TXs and the Rician factor of channels, where $N_\text{t}$ = 64, $N_\text{r}$ = 64, $L$ = 64, $\kappa$ = 0.1, the average SNR $\gamma$ = 15 dB, if $N_\text{t}$, $\kappa$ and $\gamma$ are not used as X-axis variables. It is clearly seen that all curves decline with the average SNR, the number of TXs and the Rician factor increasing. Moreover, the curves of \texttt{Theo} almost coincide with \texttt{Simu} in all modulation schemes of PSK, which indicates that the theoretical approximation is quite accurate. By comparison, the BPSK achieves the lowest ABEP and an increase in modulation order of PSK lifts the ABEP. \par
\begin{figure}
	\label{PAAbefore}
	\centering
	\subfigure[Original QR code]{
		\begin{minipage}{2.5cm}
                        \includegraphics[width=\textwidth]{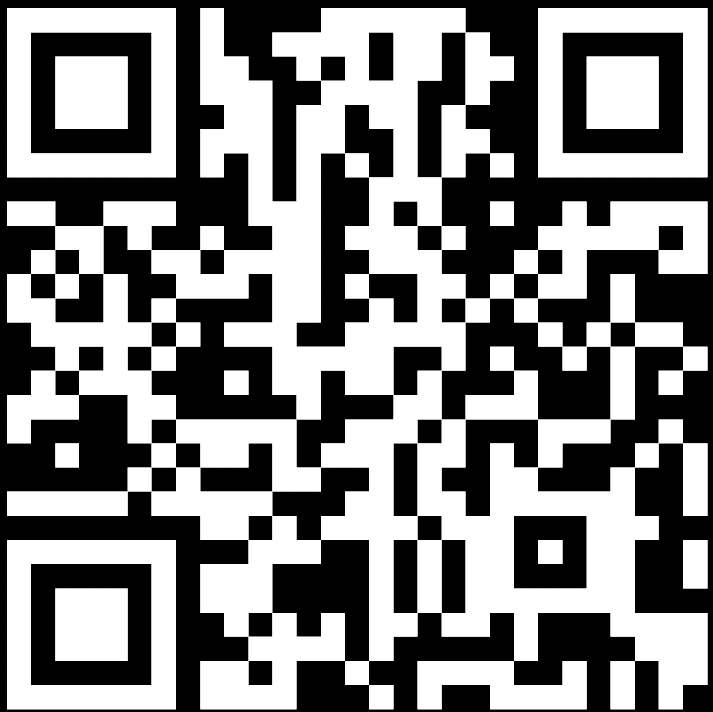} \\
		\end{minipage}
    \label{Fig7a}
	}
	\subfigure[Recovered unrecognizable QR code]{
		\begin{minipage}{2.5cm}
                        \includegraphics[width=\textwidth]{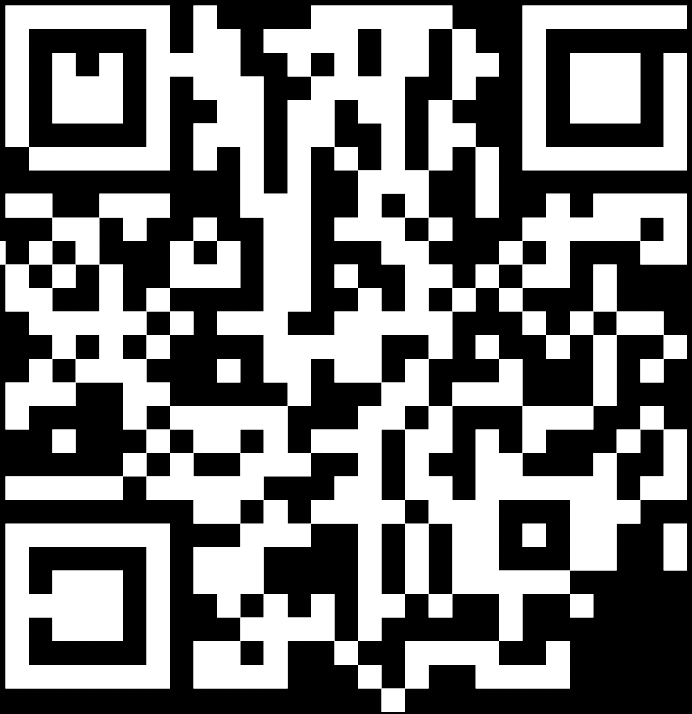} \\
		\end{minipage}
    \label{Fig7b}
	}
	\subfigure[Recovered recognizable QR code]{
		\begin{minipage}{2.5cm}
                        \includegraphics[width=\textwidth]{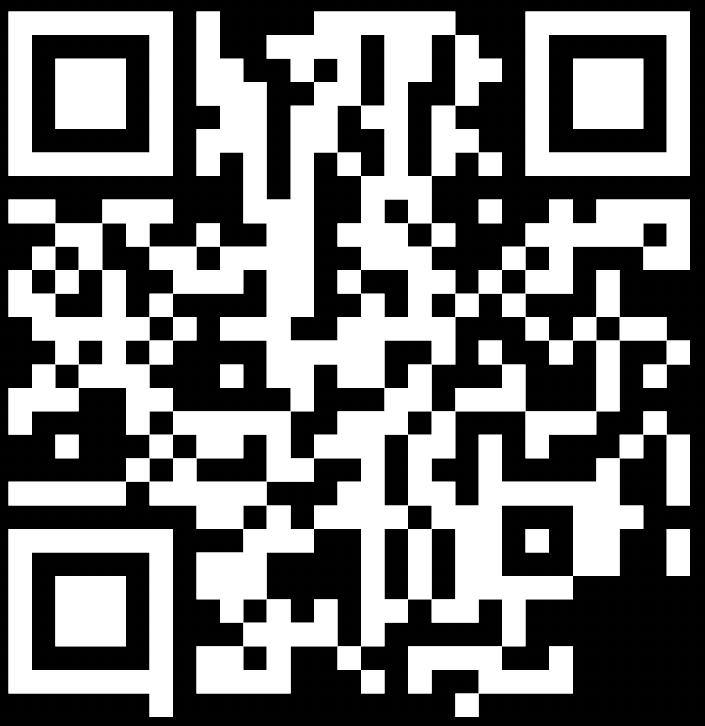} \\
		\end{minipage}
    \label{Fig7c}
	}\\
	\subfigure[Average SNR]{
		\begin{minipage}{4cm}
			\includegraphics[width=\textwidth]{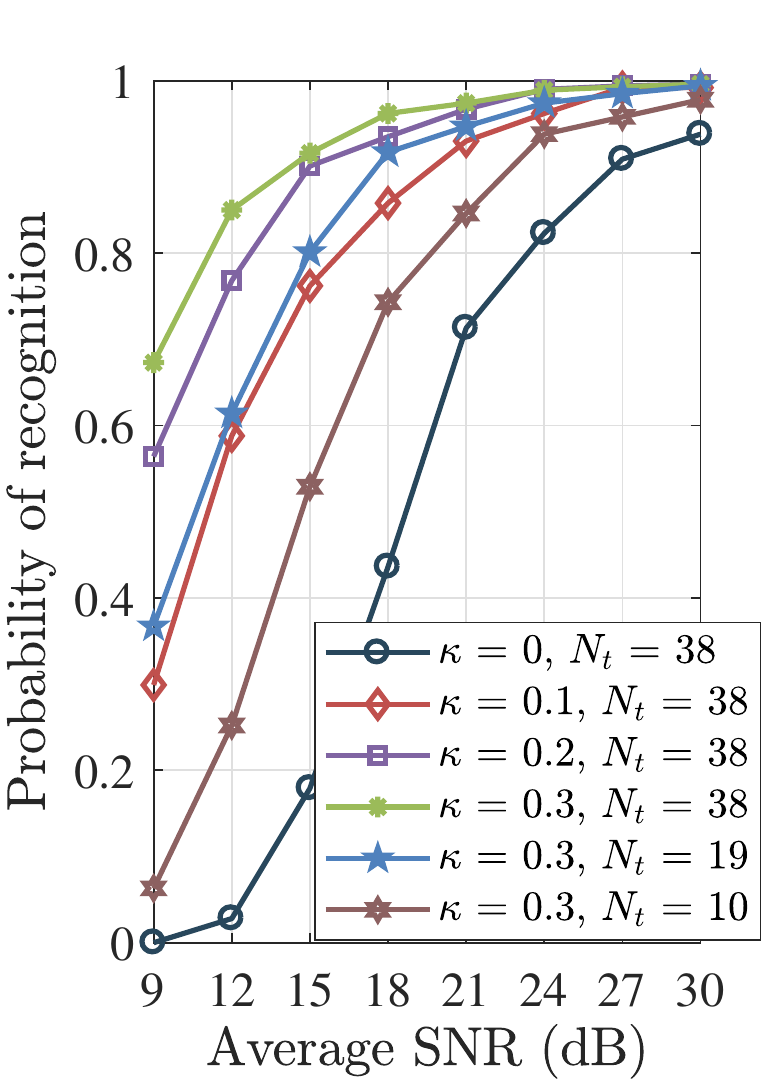} \\
		\end{minipage}
    \label{Fig7d}
	}
	\subfigure[Side length of obstruction]{
		\begin{minipage}{4cm}
			\includegraphics[width=\textwidth]{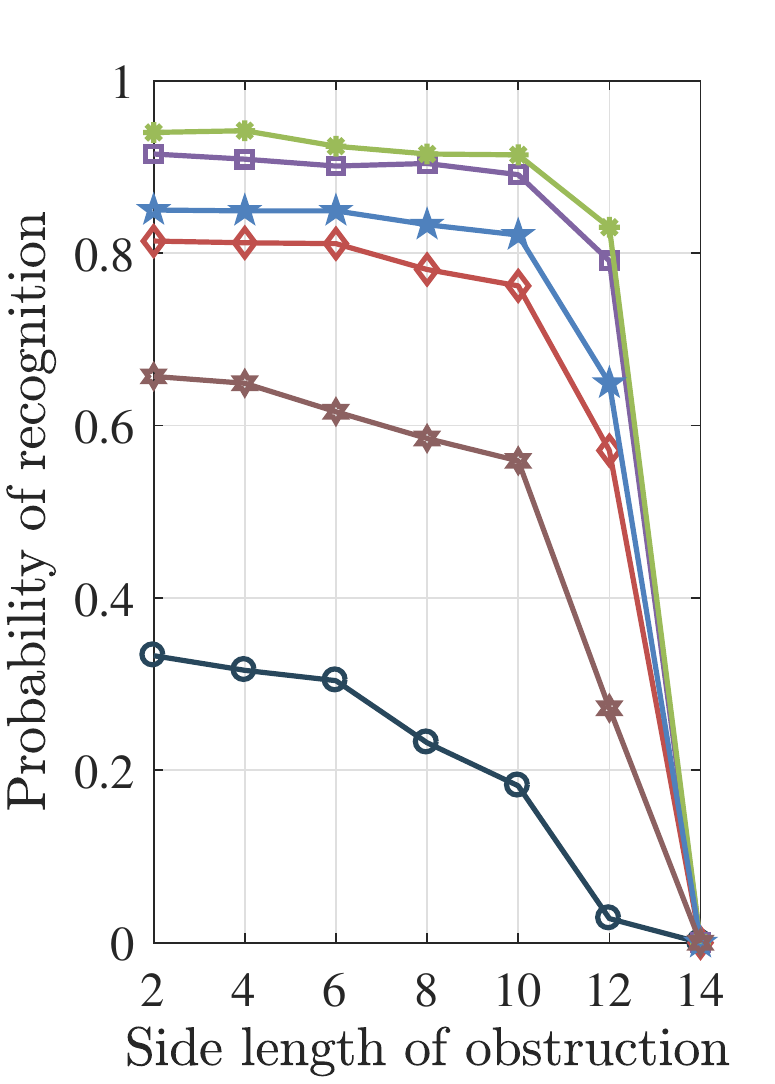} \\
			
		\end{minipage}
    \label{Fig7e}
	}
	\caption{Recovery and recognition probability of QR code for BPSK.}
    \label{Fig7}
\end{figure}
\begin{figure}
	\label{PAAbefore}
	\centering
	\subfigure[Original QR code]{
		\begin{minipage}{2.5cm}
                        \includegraphics[width=\textwidth]{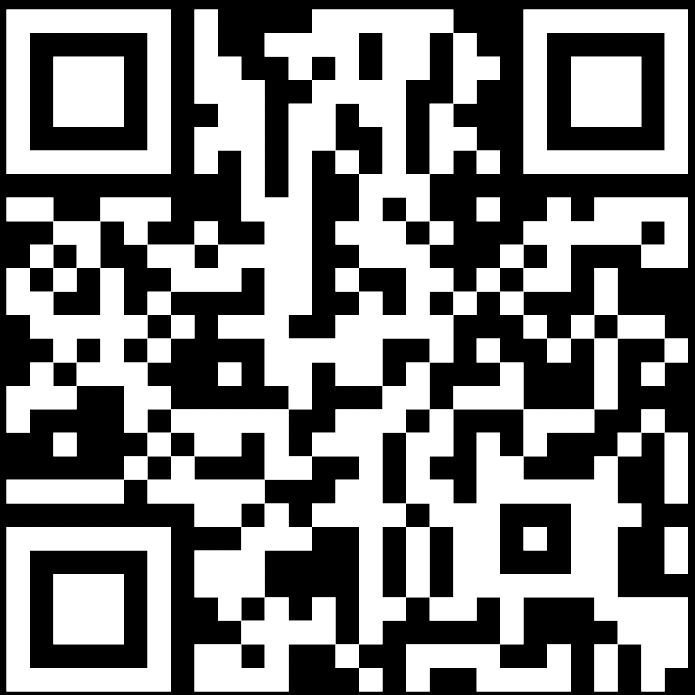} \\
		\end{minipage}
    \label{Fig8a}
	}
	\subfigure[Recovered unrecognizable QR code]{
		\begin{minipage}{2.5cm}
                        \includegraphics[width=\textwidth]{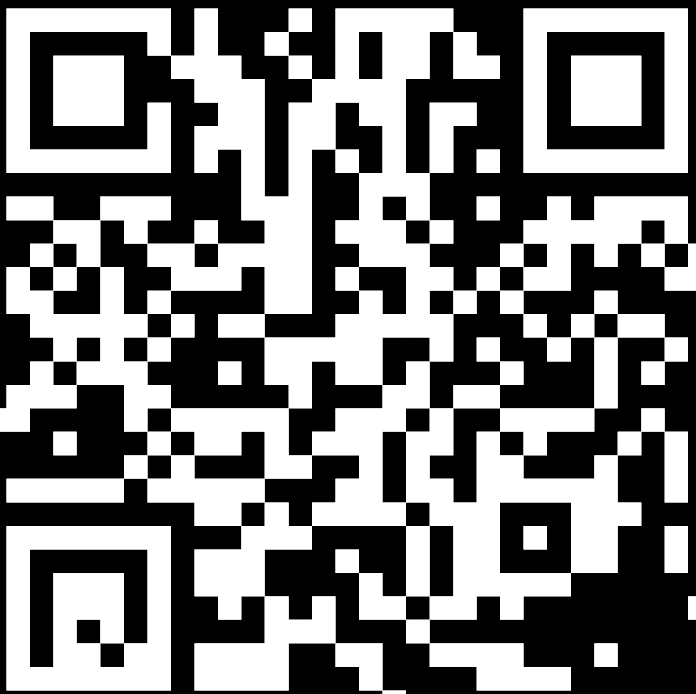} \\
		\end{minipage}
    \label{Fig8b}
	}
	\subfigure[Recovered recognizable QR code]{
		\begin{minipage}{2.5cm}
                        \includegraphics[width=\textwidth]{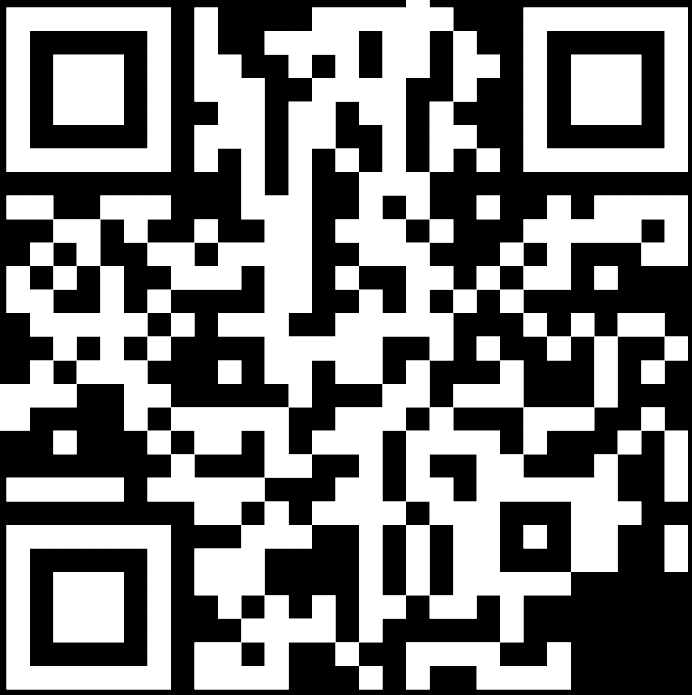} \\
		\end{minipage}
    \label{Fig8c}
	}\\
	\subfigure[Average SNR]{
		\begin{minipage}{4cm}
			\includegraphics[width=\textwidth]{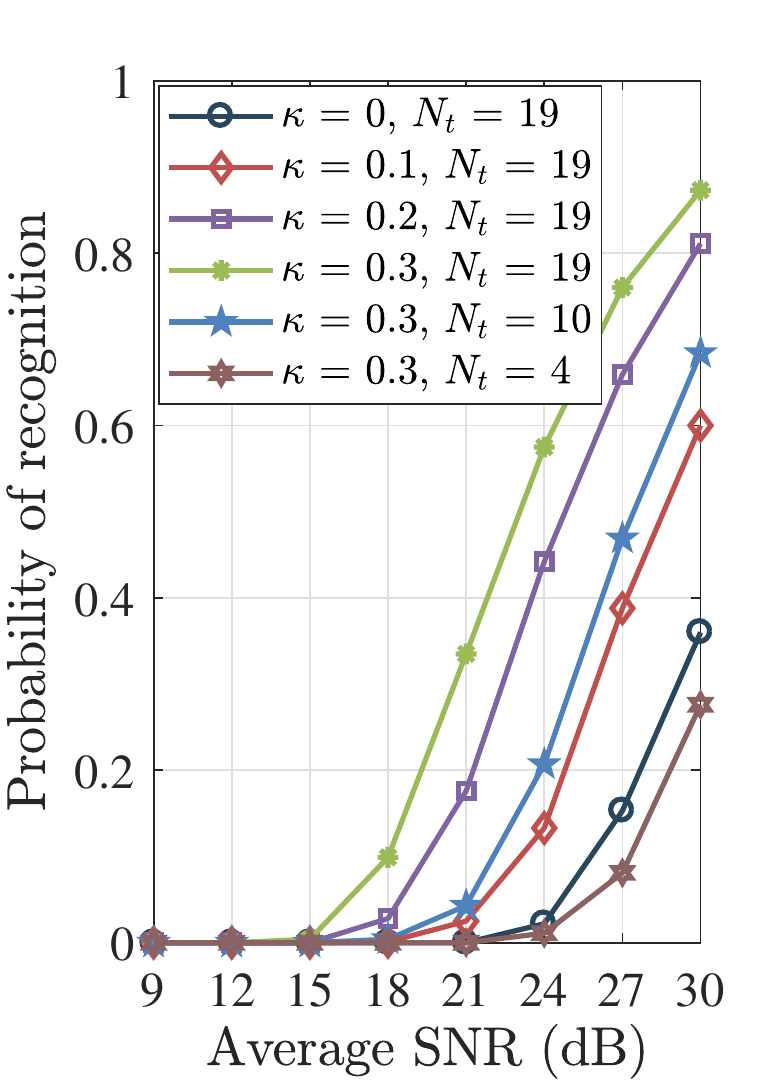} \\
		\end{minipage}
    \label{Fig8d}
	}
	\subfigure[Side length of obstruction]{
		\begin{minipage}{4cm}
			\includegraphics[width=\textwidth]{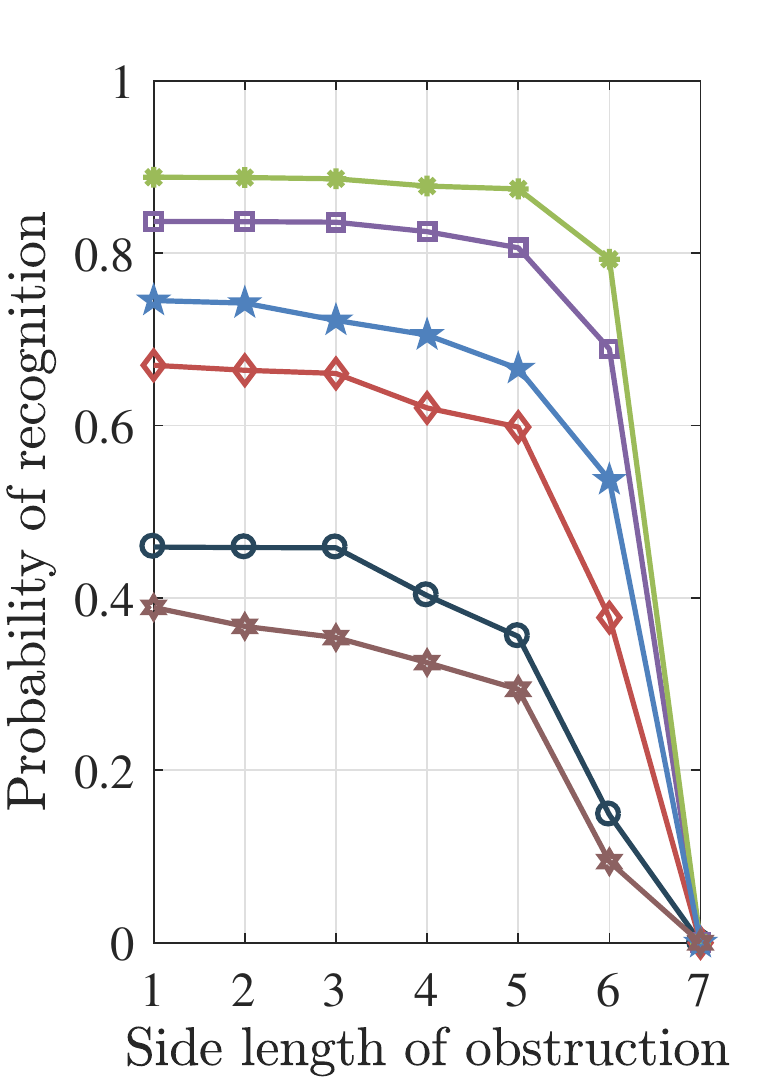} \\
			
		\end{minipage}
    \label{Fig8e}
	}
	\caption{Recovery and recognition probability of QR code for 16-PSK.}
    \label{Fig8}
\end{figure}

%
Figs. \ref{Fig7} and \ref{Fig8} show how the average SNR, the side length (measured by number of elements) of obstruction, the number of TXs and the Rician factor affect the recovery and recognition probabilities of QR code for BPSK and 16-PSK, where $N_\text{t}$ = 38, $N_\text{r}$ = 38, $L = 38^2$, $\gamma$ = 15 dB, the side length of obstruction $D$ = 10 for BPSK and $N_\text{t}$ = 19, $N_\text{r}$ = 19, $L = 19^2$, $\gamma$ = 30 dB, $D$ = 5 for 16-PSK are set. The obstruction is assumed to be square and the right bottom corner of the IRS is blocked. All QR codes shown are from a random observation for BPSK and 16-PSK. It is seen that an increase in the average SNR, the number of TXs and the Rician factor contributes to improving the recognition probability of QR code, while the side length of obstruction does the reverse.

%
%
%
%
%
%
%
%
\section{Conclusions}
This letter proposed to employ IRS-based microwave QR code for passive radio communication and investigated some fundamental agendas, involving ABEP for signal modulation, QR code implementation on an IRS, transmission design, detection, etc. According to theoretical analysis and simulation results, it can be concluded that: 1) IRS-based microwave QR code achieves a satisfactory communication performance, thereby being feasible; 2) BPSK has the lowest ABEP, and an increase in modulation order of PSK lifts the ABEP with the elements for displaying QR code reduced; 3) Higher average SNR, more TXs and larger Rician factor contribute to lowering the ABEP and probability of recognizing QR code; 4) IRS-based QR code has a good robustness to obstruction.
%
%

%
%
%
%
%
\ifCLASSOPTIONcaptionsoff
  \newpage
\fi
%


%

%
\balance
%
%




\end{document}